\documentclass[prl,twocolumn,amsmath,amssymb]{revtex4}

\usepackage{graphicx}
\usepackage{amssymb}

\newcommand{\al}{\alpha}
\newcommand{\be}{\beta}

\newcommand{\ep}{\epsilon}

\newcommand{\1}{{\bf  r}_1}
\newcommand{\2}{{\bf  r}_2}

\begin{document}
\title{Surface Criticality and Multifractality
at Localization Transitions}

\author{A. R.~Subramaniam$^1$, I. A.~Gruzberg$^1$,
A. W. W.~Ludwig$^2$, F.~Evers$^{3}$, A.~Mildenberger$^{4}$, and A.
D.~Mirlin$^{3,5,*}$ }

\affiliation{ \centerline{\hbox{$^{1}$James Franck Institute,
University of Chicago, 5640 S. Ellis Avenue, Chicago, Illinois 60637, USA}}\\
\centerline{\hbox{$^{2}$Physics Department, University of
California, Santa Barbara, California 93106, USA}}\\
$\mbox{}^{3}$ Institut f\"ur Nanotechnologie, Forschungszentrum
Karlsruhe, 76021 Karlsruhe, Germany \\
$\mbox{}^{4}$ Fakult\"at f\"ur Physik, Universit\"at Karlsruhe,
76128 Karlsruhe, Germany\\
\centerline{\hbox{$^{5}$ Institut f\"ur Theorie der kondensierten
Materie, Universit\"at Karlsruhe, 76128 Karlsruhe, Germany}}}

\date{Received 2 December 2005; published 27 March 2006}

\begin{abstract}
We develop the concept of surface multifractality for
localization-delocalization (LD) transitions in disordered
electronic systems. We point out that the critical behavior of
various observables related to wave functions near a boundary at a
LD transition is different from that in the bulk. We illustrate
this point with a calculation of boundary critical and
multifractal behavior at the 2D spin quantum Hall transition and
in a 2D metal at scales below the localization length.\\

\noindent DOI: 10.1103/PhysRevLett.96.126802 \hfill PACS numbers:
73.20.Fz, 72.15.Rn, 73.43.$-$f

\end{abstract}

\maketitle

Localization-delocalization (LD) transitions in disordered
electronic systems represent a remarkable class of quantum phase
transitions. One of their striking features is that moments of
wave functions near a LD transition are characterized by a set of
independent critical exponents \cite{Wegner80}. Criticality of the
wave functions manifests itself in their multifractality (MF)
\cite{Castellani-Peliti-86, MF-reviews}, which has been studied
analytically and numerically for a variety of systems: Anderson
transition in $d = 2 + \epsilon$, 3, and 4 dimensions
\cite{Wegner80,MF-Anderson-numerics}, as well as weak MF
\cite{MF-weak}, random Dirac fermions \cite{MF-Dirac}, integer
quantum Hall (IQH) \cite{MF-IQH}, spin quantum Hall (SQH)
\cite{Mirlin03} transitions in two dimensions, and power-law
random banded matrices \cite{MF-RBM}.

It is known that at conventional phase transitions, different
critical behavior occurs at a boundary, as compared to the bulk of
the sample \cite{critical-surface}. This is especially well
understood in two dimensions where methods of conformal field
theory (CFT) are available \cite{Cardy84}.

In this Letter we propose to extend the concept of surface
criticality to LD transitions. Such an extension has several
potentially important applications. First of all, LD transitions
are studied experimentally mainly through transport measurements,
which are performed by attaching leads to the boundary of a finite
sample. With such a setup it is feasible to measure surface
critical behavior directly. Secondly, we will show that even if
the MF of wave functions in the whole sample is studied, the
surface effects are fundamentally important. Further, LD
transitions in two dimensions are expected to be described by
certain CFTs. Such a description remains elusive for the primary
example of the IQH transition, though some proposals have been put
forward \cite{CFT-IQH-Zirnbauer,CFT-IQH-Tsvelik}. Studying MF of
wave functions near a surface is expected to help identify the CFT
for the IQH transition and similar systems.

Let us start with a brief review of bulk MF in the context of LD
transitions. The MF of wave functions $\psi(\bf r)$ at a LD
transition is characterized by the scaling of moments of
$|\psi({\bf r})|^2$ with system size $L$:
\begin{align}
L^d \langle |\psi({\bf r})|^{2q} \rangle &\sim L^{-\tau_q}, &
\tau_q &\equiv d(q-1) + \Delta_q, \label{e1b}
\end{align}
where $\langle\ldots\rangle$ denotes the disorder average. Note
than one often introduces fractal dimensions $D_q$ via
$\tau_q=D_q(q-1)$. In a metal $D_q=d$, while at a critical point
$D_q$ is a non-trivial function of $q$, implying the MF of wave
functions. Non-vanishing anomalous dimensions $\Delta_q\equiv
(q-1)(D_q-d)$ distinguish a critical point from a metallic phase
and determine the scaling of wave function correlations. Among
them, $\Delta_2 < 0$ plays the most prominent role, governing the
spatial correlations of the intensity $|\psi|^2$: $L^{2d} \langle
|\psi^2 ({\bf r}) \psi^2({\bf r}')|\rangle \sim (|{\bf r} - {\bf
r}'|/L)^{\Delta_2}$. This equation, which in technical terms
results from an operator product expansion of the field theory
\cite{duplantier91}, can be obtained from (\ref{e1b}) by using the
fact that the wave function amplitudes become essentially
uncorrelated at $|{\bf r} - {\bf r}'|\sim L$. Scaling behavior of
higher order spatial correlations, $\langle|\psi^{2q_1}({\bf r}_1)
\psi^{2q_2}({\bf r}_2) \ldots \psi^{2q_n}({\bf r}_n)|\rangle$, can
be found in a similar way. Above, the points ${\bf r}_i$ were
assumed to lie in the bulk of a critical system. In this case we
denote the multifractal exponents by $\tau_q^{\rm b}$,
$\Delta_q^{\rm b}$, etc.

In finite electronic systems wave functions vanish on the
boundary. Therefore, to assign a meaning to the MF on the
boundary, we have to interpret surface correlation functions in
the sense of the theory of critical phenomena
\cite{critical-surface,Cardy84}. Specifically, by ``points on the
boundary'' we mean points that lie close to it, as compared to
other distances, e.g. a few lattice spacings away.

With this caveat in mind, we now generalize the notion of MF to
systems with boundaries. First of all, we note that in general
even the average value of $|\psi({\bf r})|^2$ for $\bf r$ at the
boundary may be non-trivial, $\langle|\psi({\bf r})|^2\rangle
\propto L^{-d-\mu}$. In particular, $\mu$ can be non-zero in
systems with unconventional (chiral or Bogoliubov-de Gennes)
symmetry \cite{classification} which are attracting particular
interest in connection with physics of disordered superconductors.
For the moments of the intensity, we get in analogy with
Eq.~(\ref{e1b}),
\begin{eqnarray}\label{tau-q-s}
&& L^{d-1} \langle |\psi({\bf r})|^{2q} \rangle \sim L^{-\tau_q^{\rm
s}}, \\
&& \tau_q^{\rm s} = d(q-1) + q\mu + 1 + \Delta_q^{\rm s},
\label{delta-q-s}
\end{eqnarray}
with a new set of {\it surface multifractal exponents\/}
$\tau_q^{\rm s}$, $\Delta_q^{\rm s}$, which are in general independent
on their bulk counterparts \cite{note-dqs}. The normalization factor
$L^{d-1}$ is chosen such that Eq.~(\ref{tau-q-s}) yields the
contribution of the surface to the inverse participation ratios
$\langle P_q\rangle = \langle \int d^d {\bf r} |\psi({\bf
r})|^{2q} \rangle $ conventionally studied in the framework of the
MF analysis. The exponents $\Delta_q^{\rm s}$ as defined in
Eq.~(\ref{delta-q-s})  vanish in a metal and
govern statistical fluctuations of wave
functions at the boundary, $\langle |\psi({\bf r})|^{2q} \rangle /
\langle |\psi({\bf r})|^2 \rangle^q \sim L^{-\Delta_q^{\rm s}}$,
as well as their spatial correlations, e.g. $L^{2(d+\mu)} \langle
|\psi^2({\bf r})\psi^2({\bf r}')|\rangle \sim (|{\bf r} - {\bf
r}'|/L)^{\Delta_2^{\rm s}}$.

In fact, these notions can be further generalized. First, for
multi-point correlation functions, some points may be in the bulk
while the rest on the surface. Then the scaling behavior will be
described by mixed bulk-surface exponents. Second, surface and
mixed exponents will sensitively depend on the global geometry of
the boundary. For example, if the point $\bf r$ in
Eq.~(\ref{tau-q-s}) lies near the edge of a wedge with the opening
angle $\theta$, multifractal exponents will continuously depend on
$\theta$. We relegate the analysis of these generalizations to a
future publication \cite{long paper} and concentrate here on the
fundamental surface exponents $\tau_q^{\rm s}$, $\Delta_q^{\rm
s}$.

We now illustrate these points in the case of the 2D SQH plateau
transition, which belongs to symmetry class C (in the
classification of Ref.~\cite{classification}), relevant for the
description of quasiparticle transport in singlet superconductors
with broken time-reversal symmetry \cite{SQHE}. A remarkable
feature of the SQH effect is that a number of basic observables
can be calculated exactly, as was discovered in
Refs.~\cite{percolation-map,Beamond2002}, via a mapping of the
corresponding network model \cite{kagalovsky} (similar to the
IQH-network of Ref.~\cite{Chalker88}) to the problem of classical
percolation. This mapping was extended in Ref.~\cite{Mirlin03} to
extract analytical values of the bulk exponents $\Delta_2^{\rm b}
= -1/4$, $\Delta_3^{\rm b} = -3/4$.

In what follows, we generalize this  mapping to the SQH network
with a boundary, and use it to extract new surface critical and
multifractal exponents. The SQH network consists of directed links
$r$ that carry doublets of complex fluxes $\psi_\alpha(r)$
representing propagation of spin $1/2$ particles. Effects of
disorder are introduced through random SU(2) scattering matrices
on the links. At each node the scattering from two ingoing to two
outgoing links is described by the (spin-independent) matrix $S$,
with $S_{11} = S_{22} = (1-t^2)^{1/2}$, $S_{12} = -S_{21} = t$.
The value $t=1/\sqrt{2}$ corresponds to the critical point of the
SQH transition.

In order to study the effect of boundaries, we impose reflecting
boundary conditions along one direction. This does not affect the
symmetry of the system, which enables us to retain the mapping to
percolation. In technical terms, within the approach of
Ref.~\cite{percolation-map}, supersymmetry (SUSY) is preserved at
the boundary nodes of the network model. We have also checked
\cite{long paper} that the SUSY method can be extended to the two-
and three-point functions needed to extract $\tau_q^{\rm s}$ with
$q=2,3$, and yields results identical with the path integral
approach developed in Refs.~\cite{Beamond2002,Mirlin03}.

We calculate first the average local density of states (LDOS) at a
point ${\bf r}_1$ on the boundary that will allow us to find the
average of the intensity $|\psi_\alpha({\bf r}_1)|^2$. The LDOS
can be expressed in terms of one-point Green's functions and
becomes, when mapped to percolation,
\begin{equation}
\langle \rho(\1,\ep) \rangle = (1/2\pi) \bigl[1-\sum_{N}P(\1;N)
\cos{2N\ep}\bigr],
\label{ldos}
\end{equation}
where $\epsilon$ is the energy and $P(\1;N)$ is the probability of
a $N$-link hull passing through $\1$, in analogy with
\cite{Beamond2002}. The corresponding surface critical exponent is
$x_1^{\rm s} = 1/3$ \cite{surface exponent}, which should be
contrasted to its bulk value $x_1^{\rm b} = 1/4$. The latter value
implies that the percolation hull has fractal dimension $2 -
x_1^{\rm b} = 7/4$, so that $P({\bf r},N)\sim N^{-8/7}$ for ${\bf
r}$ in the  bulk. This yields, according to Eq.~(\ref{ldos}), the
DOS scaling $\rho(\epsilon)\propto \epsilon^{x_1^{\rm
b}/(2-x_1^{\rm b})} = \epsilon^{1/7}$
\cite{percolation-map,Beamond2002}. Note that $2-x_1^{\rm b} =
7/4$ is the dynamic exponent governing the scaling of energy with
the system size $L$ at SQH criticality, so that the level spacing
at $\epsilon=0$ (and thus the characteristic energy of critical
states) is $\delta\sim L^{-7/4}$.

In our case, when the point ${\bf r}_1$ is located at the surface, we
find $P({\bf r}_1,N)\sim N^{-1-x_1^{\rm s}/(2-x_1^{\rm b})} =
N^{-25/21}$ and the LDOS scaling $\rho({\bf r}_1,\epsilon)\propto
\epsilon^{x_1^{\rm s}/(2-x_1^{\rm b})} = \epsilon^{4/21}$. For the
wave function at the boundary, this implies
\begin{equation}\label{intensity}
L^2\langle|\psi_\alpha({\bf r}_1)|^2\rangle \sim L^{x_1^{\rm b} -
x_1^{\rm s}} = L^{-1/12}.
\end{equation}
Therefore, the average intensity of a critical wave function is
suppressed at the boundary with the exponent $\mu = 1/12$. A
similar calculation for the conductance between two point contacts
${\bf r}_1$ and ${\bf r}_2$ located at the boundary
\cite{note-two-terminals} yields scaling
with exponent $ 2x_1^{\rm s}=2/3$,
\begin{equation}\label{conductance}
g^{s}_{\rm pt} (\1,\2) \propto |\1 - \2|^{-2/3}.
\end{equation}

We turn now to the multifractal exponents. To calculate
$\Delta_2^{\rm s}$, we consider the correlation function
\cite{Mirlin03}:
\begin{align}
(2\pi)^{-2} \tilde{\mathcal D}(\1,\2, \ep_1, \ep_2) &= \Bigl
\langle \sum_{ij\al\be} |\psi_{i\al}(\1)| ^2 |\psi_{j\be}(\2)|^2
\nonumber \\
& \quad \times \delta(\ep_1-\ep_i) \delta(\ep_2-\ep_j) \Bigr
\rangle. \label{wave-func-corr}
\end{align}
To study critical states, we take $\epsilon_{1,2}=0$ and broaden
delta-functions by $\delta$. Using the mapping to percolation, we
relate (\ref{wave-func-corr}) to percolation probabilities,
\begin{align} \label{Dtilde}
& \tilde{\mathcal D}(\1,\2,z) = 4\sum_N \big(1-z^{2N}\big)
P(\1,\2;N) \nonumber \\
& \quad + 4\sum_{N,N'} \big(1-z^{2N}\big) \big(1-z^{2N'}\big)
P_-(\1,\2;N,N'),
\end{align}
in analogy with Ref.~\cite{Mirlin03}. Here $z=e^{-\delta}$,
$P(\1,\2;N)$ is the probability of an $N$-link hull passing
through $\1,\2$, and $P_-(\1,\2;N,N')$ is the probability of an
$N$-link hull passing through $\1$ and a different
$N^{\prime}$-link hull passing through $\2$. In view of the
cancellation of leading order terms at $z \rightarrow 1$, we need
to consider the next, sub-leading term. Using again the surface
critical exponent $x_1^{\rm s} = 1/3$, we find, when both ${\bf
r}_1$ and ${\bf r}_2$ lie at the boundary
$$
P(\1,\2;N) \sim N^{-25/21} r^{-1/3}, \qquad
r\equiv|{\bf r}_1-{\bf r}_2|\lesssim N^{4/7}.
$$
Substituted in Eq. (\ref{Dtilde}), this gives
the value
\begin{equation} \label{delta2}
\Delta^s_2 = - x_1^{\rm s} = - 1/3.
\end{equation}
A similar calculation can be done for the three-point function, with
the result
\begin{equation} \label{delta3}
\Delta_3^{\rm s} = - 3 x_1^{\rm s} = - 1.
\end{equation}
In this case there is no cancellation in the limit $z \rightarrow
1$, and the result conforms with the usual scaling of a
three-point function at the boundary.

\begin{figure}[t]
\begin{center}
\includegraphics[width=0.8\columnwidth]{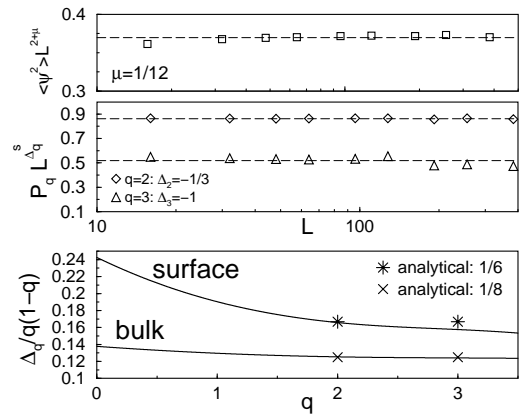}
\vspace*{-0.3cm} \caption{Surface MF at the SQH transition. The
numerical data confirm analytically found exponents (dashed
lines), with deviations as low as $4\%$ ($\mu$), $0.5\%$
($\Delta_2^{\rm s}$) and $6\%$ ($\Delta_3^{\rm s}$). Lower panel:
numerical results for the surface MF spectrum $\Delta_q^s$ for
$0<q<3.5$. For comparison, the bulk spectrum $\Delta_q^{\rm b}$ is
also shown, as well as the analytical results for $q=2,3$. }
\label{sqhe-tauq}
\end{center}
\vspace*{-0.9cm}
\end{figure}

We have verified these predictions by direct simulation of the SQH
network. To this end we have numerically diagonalized \cite{num1}
the $4L^2\times 4L^2$ discrete time-evolution operator of the
network model and selected, for each realization of disorder, two
eigenstates $\psi_{i\alpha}(r)$ with lowest quasi-energies
$\epsilon_i$. We found $\Delta_{q}^{s}$ from the scaling of the
moments $\langle|\psi_{i\alpha}^{2q}(r)|\rangle$ averaged over the
ensemble of $10^4 - 10^5$ systems. As shown in
Fig.~\ref{sqhe-tauq}, the numerical data fully confirm the values
of the exponents $\mu$, $\Delta_2^{\rm s}$, and $\Delta_3^{\rm s}$
obtained above, Eqs.~(\ref{intensity}), (\ref{delta2}), and
(\ref{delta3}). Clearly, numerical simulations allow us to study
also the scaling away from the analytically accessible values
$q=2,3$. In the lower panel, we show the results for the exponents
$\Delta_q^{\rm s}$ [divided by $q(1-q)$] in the range of $q$
between 0 and 3.5 and demonstrate their independence from the bulk
exponents $\Delta_q^{\rm b}$.

\begin{figure}[b]
\begin{center}
\includegraphics[width=0.7\columnwidth]{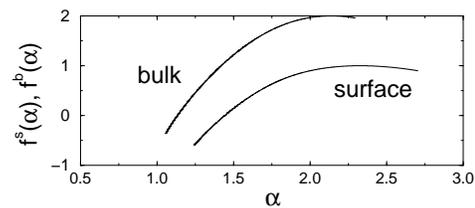}
\vspace*{-0.5cm} \caption{Surface and bulk $f(\alpha)$ spectra at
the SQH transition obtained from the data of
Fig.~\ref{sqhe-tauq}.} \label{sqhe-fa}
\end{center}
\vspace*{-0.7cm}
\end{figure}

We also show in Fig.~\ref{sqhe-fa} the singularity spectrum
$f(\alpha)$, obtained from  $\tau(q)$ by Legendre transformation.
Its meaning is as follows: the number of points $r$ in the sample
where the wave function intensity is $|\psi^2(r)| \sim
L^{-\alpha}$ scales as $L^{f(\alpha)}$. The difference between the
maximal values (2 vs. 1) of $f^{\rm b}(\alpha)$ and $f^{\rm
s}(\alpha)$ simply reflects the different dimensionalities of the
bulk and the surface. On the other hand, the difference in the
position of the maximum ($\alpha_0^{\rm b}\simeq 2.137$ vs.
$\alpha_0^{\rm s}\simeq 2.326$) and in the width of the curve
demonstrates that at the boundary the typical value of the
intensity is suppressed, while  fluctuations are stronger than in
the bulk.

We have thus shown that surface MF differs significantly from that
in the bulk. One can now ask the following question. Imagine that
one performs a multifractal analysis for the whole sample, without
separating it into ``bulk'' and ``surface''. Would then the
surface exponents play any role? A naive answer is no: since the
weight of surface points is down by a factor $1/L$, one could
expect that only the bulk exponents would matter. This is not
true, however. To illustrate this point, we turn to an example,
where the whole multifractal spectrum can be studied analytically.
Specifically, we will consider a 2D weakly localized metallic
system (dimensionless conductance $g \gg 1$), which shows weak MF
\cite{MF-weak} on length scales below the localization length
$\xi\sim e^{(\pi g)^\beta}$, where $\beta = 1$ (2) for systems
with preserved (broken) time-reversal symmetry. With minor
modifications, the formulas below describe also the Anderson
transition in $2 + \epsilon$ dimensions.

\begin{figure}
\begin{center}
\includegraphics[width=0.7\columnwidth]{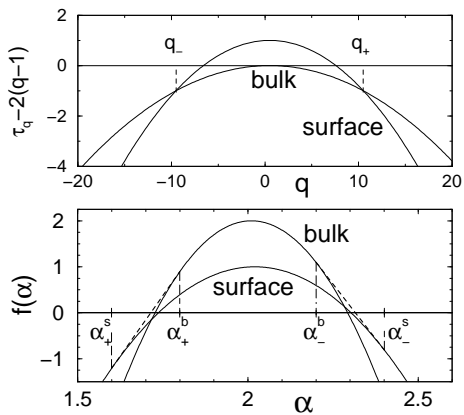}
\vspace*{-0.3cm} \caption{Surface and bulk multifractal spectra
$\tau_q$ and $f(\alpha)$ for a 2D metal with $\gamma=0.01$. For
details see text. } \label{fig-2d}
\end{center}
\vspace*{-1cm}
\end{figure}

The bulk multifractal spectrum of this system was obtained via the
perturbative renormalization group treatment of the underlying field
theory ($\sigma$-model) \cite{Wegner80} and also within the
instanton approach \cite{MF-weak}. The result reads
\begin{equation}\label{2d-bulk-tauq}
\tau_q^{\rm b} = 2(q-1) + \gamma q (1-q); \qquad \gamma =
(\beta\pi g)^{-1} \ll 1.
\end{equation}
Generalizing this calculation to the surface case, we find
\begin{equation}\label{2d-surface-tauq}
\tau_q^{\rm s} = 2(q-1) + 1 + 2\gamma q (1-q).
\end{equation}
The factor of 2 in front of the last (anomalous) term can be
traced back to the corresponding enhancement of the return
probability near the surface. Performing the Legendre
transformation, we find the $f(\alpha)$-spectra,
\begin{eqnarray} \label{2d-bulk-fa}
f^{\rm b}(\alpha) &=& 2 - (\alpha-2-\gamma)^2/4\gamma , \\
f^{\rm s}(\alpha) &=& 1 - (\alpha-2-2\gamma)^2/8\gamma .
\label{2d-surface-fa}
\end{eqnarray}
These results are illustrated in Fig.~\ref{fig-2d}. When the MF in
the whole sample is analyzed, the lowest of the $\tau_q$ exponents
``wins''. It is easy to see that the surface effects become
dominant outside the range $q_- < q < q_+$, where $q_\pm \simeq
\pm \gamma^{-1/2}$ are the roots of the equation $\tau_q^{\rm
b}=\tau_q^{\rm s}$. The lower panel of Fig.~\ref{fig-2d} shows how
this is translated into the $f(\alpha)$ representation. The total
singularity spectrum is given by the bulk function $f^{\rm
b}(\alpha)$ only for $\alpha_+^{\rm b} < \alpha < \alpha_-^{\rm
b}$, where $\alpha_\pm^{\rm b}-2\simeq \mp 2 \gamma^{1/2}$.
Outside this range the surface effects are important.
Specifically, $f(\alpha)$ is equal to the surface spectrum $f^{\rm
s}(\alpha)$ for $\alpha < \alpha_+^{\rm s}$ and $\alpha >
\alpha_-^{\rm s}$, where $\alpha_\pm^{\rm s}-2\simeq \mp 4
\gamma^{1/2}$, while in the intermediate intervals $\alpha_+^{\rm
s} < \alpha < \alpha_+^{\rm b}$ and $\alpha_-^{\rm b} < \alpha <
\alpha_-^{\rm s}$ its dependence on $\alpha$ becomes linear (shown
by dashed lines). The latter behavior is governed by intermediate
(between  ``bulk'' and ``surface'') points with a distance from
the surface $r\sim L^\beta$, $0 < \beta < 1$; their $f(\alpha)$
spectrum is easily found to be $f_\beta(\alpha)=\beta f^{\rm
b}(\alpha) + (1-\beta) f^{\rm s}(\alpha)$. Note that in this case
the surface effects modify $f(\alpha)$ in the whole range below
$f(\alpha)\simeq 1$. Therefore, the surface exponents affect the
multifractal spectrum of the sample not only for rare realizations
of disorder (governing the negative part of $f(\alpha)$) but also
in a typical sample.

To summarize, we have developed the concept of surface MF for
localization transitions in disordered electronic systems, and
have extended the notion of surface criticality to those
transitions. We have calculated surface critical and multifractal
exponents for the SQH transition. Considering the example of a 2D
weakly localized system, we have shown that the surface
criticality may crucially affect multifractal spectra in systems
with boundaries. Our work opens a new direction of research in the
field of Anderson and quantum Hall transitions. More generally, it
is interesting to study boundary effects for MF in other
stochastic systems.

We thank N. Read for initial discussions on boundary MF at the SQH
transition. This work was supported by NSF DMR-0448820, NSF MRSEC
DMR-0213745, the Alfred P. Sloan Foundation and the Research
Corporation (I. A. G.), NSF DMR-00-75064 (A. W. W. L.), U.S. DOE
OS No. W-31-109-ENG-38 (A. D. M.), SPP ``Quanten-Hall-Systeme''
and CFN of the DFG (F. E., A. M., A. D. M.).


\begin{thebibliography}{99}
\bibitem{note-dqs} In analogy with the bulk, one can consider
$L^{1+\mu}|\psi^2({\bf r})|$ as a measure on the boundary
and define the corresponding fractal dimensions $D_q^{\rm s}$ related
to $\Delta_q^{\rm s}$ from Eq.~(\ref{delta-q-s}) as
$\Delta_q^{\rm s} = (q-1)(D_q^{\rm s}-d+1)$.
\end{thebibliography}
\end{document}